\documentclass[twocolumn,showpacs,floatfix,superscriptaddress,amsmath,amssymb,prl]{revtex4}
\usepackage{graphicx}
\usepackage{amsmath}
\usepackage{dcolumn}
\usepackage{bm}

\begin{document}

\title{Measuring geometric phases of scattering states in nanoscale 
       electronic devices}

\author{Huan-Qiang Zhou}
 \email{hqz@maths.uq.edu.au}
\affiliation{Centre for Mathematical Physics, University of Queensland,
                     Brisbane Qld 4072, Australia}
\author{Urban Lundin}
\author{Sam Young Cho}
\author{Ross H.\ McKenzie}
\affiliation{Department of Physics, University of Queensland,
             Brisbane Qld 4072, Australia}

\begin{abstract}
We show how a new quantum property, a geometric phase, 
associated with scattering states can be exhibited in nanoscale 
electronic devices. 
We propose an experiment to use interference 
to directly measure the effect of the new geometric phase. 
The setup involves a double path interferometer, adapted from
that used to measure the phase evolution of electrons as they traverse a
quantum dot (QD). Gate voltages on the QD could be varied cyclically and
adiabatically, in a manner similar to that used to observe
quantum adiabatic charge pumping. The interference due to the geometric phase 
results in oscillations in the current collected in the drain
when a small bias across the device is applied. 
We illustrate the effect with examples of geometric phases resulting from
both Abelian and non-Abelian gauge potentials.
\end{abstract}
\maketitle

Nanoscale electronic devices can exhibit distinct quantum features such as
interference~\cite{schuster,ji}, entanglement~\cite{bena},
discrete charge~\cite{kastner}, the Aharonov-Bohm effect~\cite{yacoby},
and Berry's phase~\cite{berry}.
The effect of Berry's phases associated with both Abelian and non-Abelian
gauge potentials has found possible applications in quantum
computation~\cite{jones,zanardi}. 
In systems with discrete energy levels, Berry's phase makes use of the 
adiabatic theorem~\cite{schiff} and requires that the frequency of 
variation of parameters be much less than the energy level spacing. 
Berry's phase has been demonstrated
in a variety of microscopic~\cite{sw89} as
well as mesoscopic systems~\cite{fazio}.  

A natural question arises as to whether or not 
there is a geometric phase
accompanying a scattering state in a cyclic and adiabatic variation 
of external parameters 
which characterize an {\it open} system with a continuous energy spectrum. 
An important example of such scattering states are those present in a 
nanoscale electronic device coupled to electrical leads. 
This question has been addressed recently 
in the context of quantum adiabatic pumping of charge and spin in
nanoscale electronic devices~\cite{zhou1}. The latter is subject to
intense study~\cite{b98}, 
motivated by the experimental realization 
reported in the works of Marcus and co-workers~\cite{switkes,marcus}.  
It was found that quantum adiabatic scattering provides another setting 
in which both Abelian and non-Abelian gauge potentials 
arise naturally. 
It was noticed that two gauge potentials may be defined 
in terms of the row and column vectors of instantaneous (frozen) 
scattering matrix, respectively.
They are connected with each other via a time-reversal operation.
Indeed, the scattering states associated with Hamiltonian
accumulate geometric phases defined by the row vectors 
whereas the scattering states associated with the time-reversed Hamiltonian
accumulate geometric phases defined by the column vectors.
The connection between the geometric phases for the time-reversed scattering
states and quantum adiabatic pumping was clarified in Ref.\cite{zhou1}.
In fact, the same non-Abelian gauge field as that found
by Moody {\it et al.}~\cite{msw86} for a diatomic molecule also appears 
in an open 
system describing the tunneling from a scanning tunneling microscopic tip 
through a single magnetic spin~\cite{zhou1}. 
However, it remains open how to experimentally observe the geometric phase
for a scattering state itself.

In this Letter, we describe the general theory characterizing geometric 
phases for scattering states associated to a Hamiltonian with a continuous 
energy spectrum. A possible experimental setup utilizing 
nanoscale electronic devices is
proposed to directly measure the effect 
of the geometric phases in an interference experiment.
The experimental setup is similar to
that used to measure the phase evolution of electrons as they traverse a
QD, with some adaptation to accommodate
the adiabatic variation of external parameters, e.g., gate voltages.
It turns out that the geometric phase manifests itself
in oscillations in the current collected in the drain
when a small bias across the device is applied.

Consider an open quantum mechanical system characterized by the 
Hamiltonian $H(t)$ with a continuous energy spectrum, which
undergoes an adiabatic evolution. By ``adiabatic'' we mean that the time
particles ``dwell'' inside the scattering region is much shorter than the
adiabatic period. Then the system is well described by the {\it frozen}
instantaneous scattering matrix $S(t)$~\cite{thirring,avron}, 
which is a $2N \times 2N$ matrix, with 
$N$ the number of channels (such as spin) for the 
incoming and outgoing waves. Define vectors 
${\mathbf n}_\alpha = (S_{\alpha 1}, \cdots, 
 S_{\alpha, 2N}) (\alpha = 1, \cdots, N)$
in terms of the rows of the scattering matrix. These vectors are
orthonormal and so constitutes a smooth set of local bases.  
As the system undergoes an adiabatic and cyclic evolution and returns to
the initial configuration, the interplay between the adiabatic (dynamic)
evolution and the global geometric property implies that
the row vectors ${\bf n}_\alpha$ acquire a geometric
phase,
\begin{equation}
U = {\rm P} \exp (i \oint \sum_{\nu} A_\nu d V_\nu ), 
\label{gp}
\end{equation}
where P denotes path ordering, $A_{\alpha \beta \nu} 
=i\; {\bf n}^*_\beta \cdot \partial_{\nu} {\bf n}_\alpha$ 
($\partial_{\nu}\equiv \partial / \partial V_{\nu}$) 
is the gauge potential,
and $V_\nu$ are independent slowly varying
external parameters. Here we emphasize that, unlike 
Berry's phases, the causality condition plays an essential role,
which states that scattered waves appear {\it only} after
the incident wave hits the scatterer. Under the gauge transformation
which mixes up scattering states from different channels
$ {\bf n}'_\alpha = \sum _\beta \omega _{\alpha \beta} {\bf n}_\beta$, 
the gauge potential defined by $A = \sum_\nu A_\nu dV_\nu$
transforms as
\begin {equation}
A' =  i d \omega \omega ^{-1} + \omega A \omega ^{-1}. \label{gt}
\end{equation}
That is, $A$ describes $U(N)$ gauge potentials arising from the 
superposition of different channel scattering states.
As a special case, 
the Abelian gauge group $U(1)$ originates from the fact that the absolute 
phase is not observable in quantum mechanics. The adiabatic variation 
of the scattering potential 
$V(x,t)$ induces a local gauge transformation 
${\bf n}' = \exp (i \varphi) {\bf n}$ due to the time dependence of 
the phase $\varphi$ in quantum mechanics. 

\begin{figure}[hbt]
\includegraphics[scale=0.9]{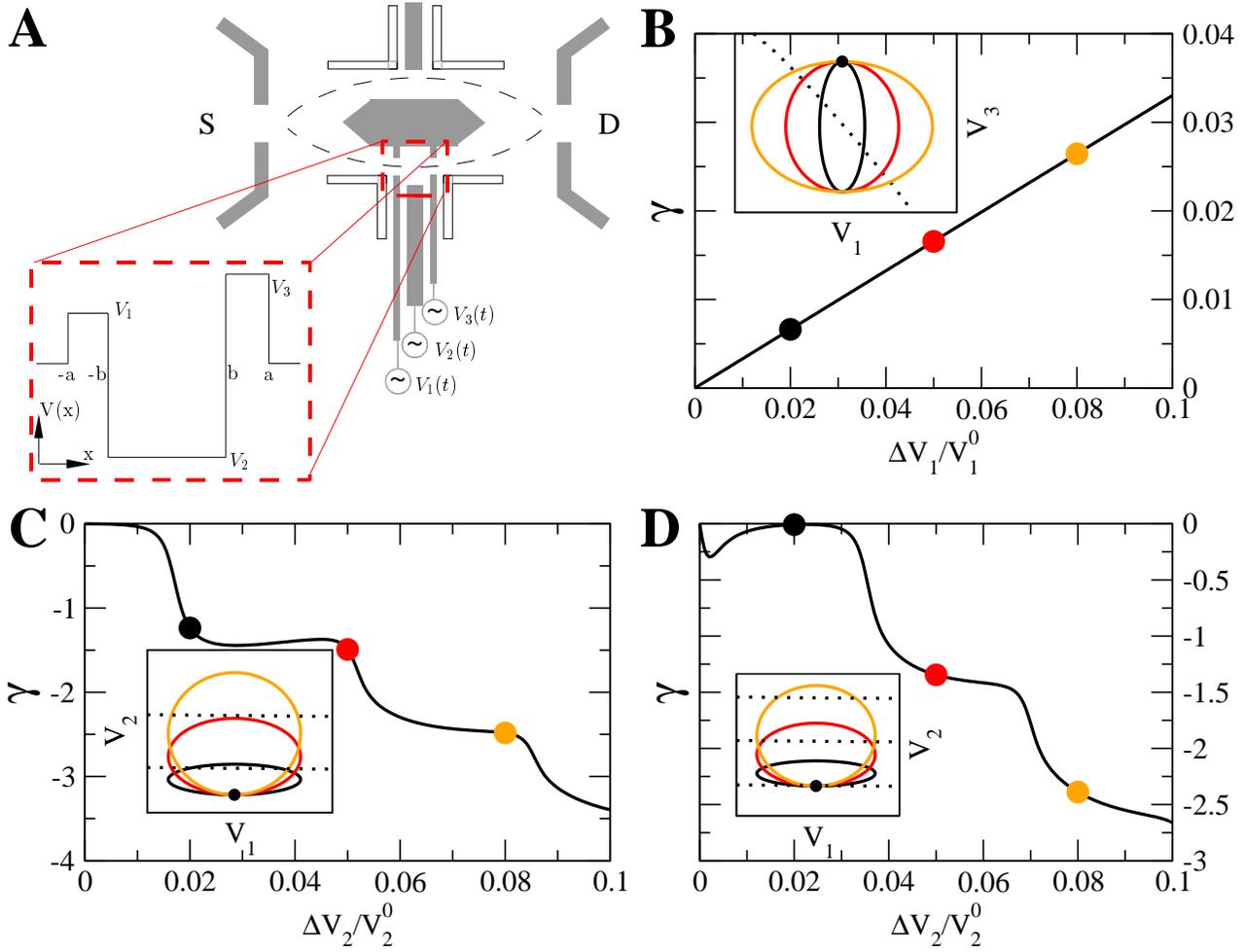}
\caption{\label{abelian:fig}
{\bf A}, Proposed electronic interferometer to directly measure
the geometric phase.
The cyclic and adiabatic variation of the gate voltages
$V_1,V_2$ and $V_3$ pairwise imposes a geometric
phase in the QD-arm. The inset shows the model one-dimensional potential inside
the QD. 
{\bf B}, Geometric phase, $\gamma$, as a function of $\Delta V_1$, with
$\Delta V_2=0$, and $\Delta V_3$ kept constant at $V^0_3/10$. 
The inset shows the contours for three different values of $\Delta V_1$,
presented as colored dots on the curve. The black dot in the inset represents
the initial state. The dotted lines in the inset show the positions of
transmission resonances through the QD in the parameter space for a
selected incident energy $E= 0.56 V^0_1$, which is off resonance.
{\bf C}, Geometric phase, $\gamma$, as a function of $\Delta V_2$, with
$\Delta V_1$ kept constant at $V^0_1/10$, and $\Delta V_3=0$. 
The inset displays the same information as in {\bf B}. 
The geometric phase, $\gamma$, shows a dip when the contour touches a
new transmission resonance. 
There is a significant change in $\gamma$ when $\Delta V_2$ varies.
{\bf D}, Same as {\bf c}, but for a different value of the incident
energy, $E$, positioned at a resonance.
Parameters (in units of $V^0_1, \hbar =m_e =1$): 
$a = 40, b=39, V^0_2=-1.0, V^0_3-1.2, \delta=\pi/2$. 
(If $V_1^0\sim 60$ mV above the Fermi level and $m^*=0.07m_e$ as for GaAs, 
then $a\sim 410$nm, and $b\sim 400$nm.}
\end{figure}

Let us now turn to a specific proposal as to how to 
experimentally observe the effect of the geometric phase, in 
a mesoscopic electronic device. We emphasize that the theory presented 
here is not restricted to mesoscopic physics, but to any system 
described by scattering states with continuous energy spectrum. 
We also emphasize that, although (for reasons of concreteness) we consider 
a specific potential for a quantum dot, the general idea applies to scattering 
states in general.  Consider a QD modeled by a potential $V(x)$ 
with $x$ denoting the coordinate 
(see Fig.~\ref{abelian:fig}A).
For reasons of simplicity, we choose the potential $V(x)$ 
as $0$ for $|x|\geq a$, $V_1$ for $-a < x < -b$, $V_2$ for $|x| \leq b$, and 
$V_3$ for $b < x < a$.  
For a QD of size 800 nm (see Fig.~\ref{abelian:fig}),
the energy level spacing is of the order of 4.5 meV. 
The Coulomb energy, assuming a dielectric constant of 10, 
is of the order of 0.08 meV. 
Thus, the dimension of the QD is such that the Coulomb energy is much
less than the separation between the resonances and can be ignored. 
Also the spin-dependent scattering inside the QD is ignored. 
Then the instantaneous $2 \times 2$ scattering matrix
$S(t)$ for the QD is determined from the solution of 
the Schr\"odinger equation
$(-(\hbar ^2 /2m) \partial ^2/ \partial x^2 + V(x)- E ) \psi =0$.
Let $r_{QD}$ and $t_{QD}$ denote, respectively, the reflection and transmission
coefficients of the QD for the left incident electron, and 
$r'_{QD}$ and $t'_{QD}$ denote, respectively, the reflection and transmission
coefficients of the QD for the right incident electron, which 
are functions of the parameters of the QD. 
If the potential is mirror symmetric, i.e., $V_1=V_3$, then $t_{QD}=t'_{QD}$ 
and $r_{QD}=r'_{QD}$ and the geometric phase is trivial. 
Therefore, to observe a nontrivial geometric phase 
it is necessary to break the mirror symmetry of the potential. 
This implies that we have to choose $V_1\neq V_3$.  

Suppose we periodically and adiabatically vary independent 
external parameters $V_1$, $V_2$, and $V_3$.
For instance, we can choose to adiabatically change $V_1$ and $V_2$ 
with $V_3$ kept constant, i.e., 
$V_1 = V^0_1 + \Delta V_1 \sin \Omega t,
~V_2 = V^0_2 + \Delta V_2 [\sin (\delta +\Omega t)-\sin \delta], V_3 = V^0_3, 
~(\Delta V_{1,2} \ll V^0_{1,2})$, 
with $\Omega$ being the slow frequency characterizing the adiabaticity
and $\delta$ the phase difference.  
(The presence of an extra term $-\Delta V_2 \sin
\delta$ is only to ensure that the initial
state is the same for all different contours.) 
In our case, 
this may be achieved by controlling the gate voltages such that
the dwell time $\tau _d$ during which electrons scatter
off the QD is much shorter than the period $T= 2 \pi /\Omega $
during which the system completes the whole adiabatic cyclic process. 
In such a limit, electrons well-defined in the incident energy $E$ 
are scattered at a 
well-defined time $t$ as measured at large time scale 
by the adiabatic cycle period $T$, consistent with the Heisenberg
uncertainty principle. That is, it makes sense to speak of the
instantaneous scattering matrix for electrons with a given incident
energy. 
Then, in addition to the dynamic phase,
the scattered waves accumulate a geometric phase factor $e^{i \gamma}$ 
during one cycle with $\gamma$ given using Eq.(\ref{gp}) for the $U(1)$ case. 
Now 
${\bf n}$ denotes the row vector of the scattering matrix $S$, i.e.,
${\bf n} = (r_{QD}, t_{QD})$, so that 
\begin{equation}
\gamma = \oint r^*_{QD} dr_{QD}+t_{QD}^*dt_{QD} = 
         \oint A_1 dV_1 + A_2 d V_2 + A_3 d V_3,
\end{equation}
since $r_{QD}$ and $t_{QD}$ depend on any variables which
vary during the cycle. 
Here we assume all $V$'s are changing with time.
However, if any of them is kept constant the corresponding term
disappear.
In this case, the gauge transformation, Eq.(\ref{gt}), becomes
\begin {equation}
A' =  A - d  \varphi. 
\end{equation}
The curvature defined by $dA$ 
is gauge invariant, which allows us to rewrite $\gamma$ in 
the form $\gamma = \iint dA$ using Stokes'
theorem, where the integral is over the area encircled by the contour.
This implies the gauge invariance of the geometric phase.
For the specific case when the variation is very small $\gamma$
is simply proportional to the area swept out in the parameter
space. The geometric phase is plotted in Fig.~\ref{abelian:fig}B 
as a function of 
$\Delta V_1$, 
with $\Delta V_2=0$, $\Delta V_3$ kept constant at $V^0_3/10$, 
 and $\delta =\pi/2$. As we
see, $\gamma$ behaves linearly as $\Delta V_1$ changes,
resulting from the fact that energy-dependent resonances on the QD are 
robust for the variation of $V_1$. However, the slope
sensitively depends on whether we are on or off resonance. 
Similarly, we plot the phase $\gamma$
in Figs.~\ref{abelian:fig}C and \ref{abelian:fig}D as 
a function of $\Delta V_2$ when $\Delta V_1$ is kept
constant at $V^0_1/10$ and $\Delta V_3=0$, 
with the incident energy $E$ being off and on a
resonance at the initial state, respectively.  The oscillating
behavior indicates that $\gamma$ is quite sensitive to the presence of the
resonant levels inside the contour
in the parameter space $\{ V_1, V_2 \}$, 
as displayed in the insets in Figs.~\ref{abelian:fig}B-D. 
A jump occurs in the geometric phase if the contour encircles a 
new transmission resonance.

Having described how the geometric phase appears for the scattering 
state using a QD, we now consider how to measure it experimentally. 
The experimental setup we propose is the double path 
interferometer (see Fig.~\ref{abelian:fig}A), which previously was used 
to measure the phase evolution of electrons 
as they traverse a QD~\cite{ji,schuster}. 
The measurement proceeds as follows. The system is prepared
in some scattering state with incident energy 
$E$ for certain initial values of the external parameters $V_1$, $V_2$ and
$V_3$, which are controllable by adjusting the Fermi level in the leads
and the attached gate voltages, respectively. Then, the gate voltages
are varied in a cyclic manner and sufficiently slowly that the system always 
remains in the instantaneous scattering state at any later instant $t$. 
Electrons in the reference path and QD-path interfere and is observed as
oscillations in the current collected in the drain 
in the linear response regime, i.e., in the presence of 
a small bias across the QD. 
A crucial feature of the device here, in contrast to the 
experimental setups used to observe adiabatic pumping 
currents~\cite{switkes}, is that the reflected electrons are 
allowed to escape from the interferometer 
between the source and the drain, thus violating current conservation.
This prevents multiple 
scattering processes which dominate quantum adiabatic pumping
as current conservation requires.  
It is this feature that makes it possible to capture the effect of the 
geometric phases for scattering states. 

The device we suggested above involves quantum interferometry of geometric
phases in a mesoscopic open system. This is similar to the Aharonov-Bohm 
effect,
which leads to an oscillating periodic component in the current as a 
function of
magnetic field applied~\cite{ji,schuster,yacoby}.  However, instead of 
the flux produced by the external magnetic field, here the 
geometric phase results from the gauge field induced by the 
adiabatic dynamics of the QD.  The total device transmission ${\cal T}$
resulting from the two-path interference after one period $T$
takes the form
\begin{equation}
{\cal T}  = |t_{ref}|^2 + |t_{QD}|^2 + 2 |t_{ref}||t_{QD}| 
\cos (\gamma + \varphi_{12}).
\label {trans}
\end{equation}
Here $t_{ref}$ denotes the transmission coefficient for the reference
path, and $\varphi _{12}$ is the phase difference between the two
transmission coefficients $t_{ref}$ and $t_{QD}$, which only depends on
the initial scattering state. In fact, Eq.(\ref{trans}) is gauge invariant, 
as it should be, and holds at any instant $t$ as long as $t_{ref}, t_{QD}$ and
$\gamma$ take the instantaneous values, because the transmission ${\cal T}$
describes the current collected in the drain and so is observable. 
However, we emphasize that {\it only} 
for the whole period $T$, is $\gamma$ gauge invariant and 
therefore observable. One may recognize that the transmission
${\cal T}$ in Eq.(\ref{trans}) 
takes the same form as that at the initial instant, except
for the involvement of the geometric phase $\gamma$ in the last term.
Indeed, the first and second terms just provide a background solely
determined by the initial state, i.e., it
does not depend on which adiabatic cycle we choose. This is in contrast to
the geometric phase $\gamma$ which does depend on contours the system
traverses in the parameter space. 
$\varphi_{12}$ also changes during the cycle, but is periodic in $T$. 
For different choices of the phase difference
$\delta$ corresponding to different shapes of the adiabatic cycles, 
the transmission
${\cal T}$ varies considerably in the entire energy range. 
>From the experimental data for the interferometer reported by 
Schuster {\it et al.}~\cite{schuster}, 
one may estimate that the background term $|t_{ref}|^2 + |t_{QD}|^2$
is approximately $1.05$ and the oscillating amplitude $2 |t_{ref}|
|t_{QD}|$ is approximately $0.05$. Hence, for such a device  
the deviation coming from 
the presence of $\gamma$ would be approximately $0.1$. 
Thus, even at a relatively low visibility the effect from the scattering 
geometric phase should be observable. 
Fig.~\ref{abelian:fig} and Eq.(\ref{trans}) imply that the effect 
of the geometric phase $\gamma$ on the transmission ${\cal T}$ is 
observable. An important issue is that the dwell time $\tau _d$ is
longer when the QD is on resonance, 
so the frequency of the adiabatic variation, $\Omega$, 
should be sufficiently slow to ensure the adiabaticity parameter
$\epsilon \equiv \Omega \tau _d$ to be very small.
We believe that current technology is sufficient to control
the adiabatic dynamics to observe the effect of the geometric phase.

Now we explain how to modify the interference setup to observe the
geometric phase associated with the true non-Abelian gauge field which
occurs in the context of adiabatic spin pumping. 
The geometric (matrix) phase 
$U$, from Eq.(\ref{gp}), is a $2 \times 2$ matrix and 
results from the true non-Abelian gauge potential, which is the 
time-reversed counterpart of that studied in quantum spin 
pumping~\cite{zhou1}. Adopting the same notations as those there, one sees
that the non-Abelian gauge potential takes the same form as Eq.(8) 
in Ref.~\cite{zhou1}, with $\phi$ replaced by $-\phi$. 
For a contour when $\phi$ varies from 0 to $2\pi$ with some
fixed $\theta$, we have
$U= \exp \{i \pi [1-\cos (\delta_1 -\delta_2)] \sin ^2 \theta \sigma ^3 
/2\}$.
In this case, the non-Abelian character of the potential is lost~\cite{zee}.
To observe the effect of the non-Abelian gauge field, it is necessary to
choose a contour which
varies both $\theta$ and $\phi$.
The noncommutativity of the matrix form 
of the gauge potential
presents some difficulties to explicitly calculate the (matrix) phase $U$. 
However,
one may use the non-Abelian version of Stokes' theorem~\cite{bralic} to 
evaluate $U$. Alternatively, in numerical calculations,
we can perform a straightforward expansion of the path ordered 
exponential, Eq.(\ref{gp}). 
The 
effect of the geometric phase is seen from the gauge invariant transmission 
\begin{equation}
{\cal T} = {\rm Tr} | t_{ref} + U t_{SI}|^2,
\label{T:eq}
\end{equation}
with $t_{ref}$ and $t_{SI}$ being the $2 \times 2$ 
transmission coefficient matrices for the reference and spin-dependent 
interaction paths, respectively. 
The interference pattern of the two 
paths is changed due to the geometric phase. 

\begin{figure}[hbt]
\includegraphics[scale=1.0]{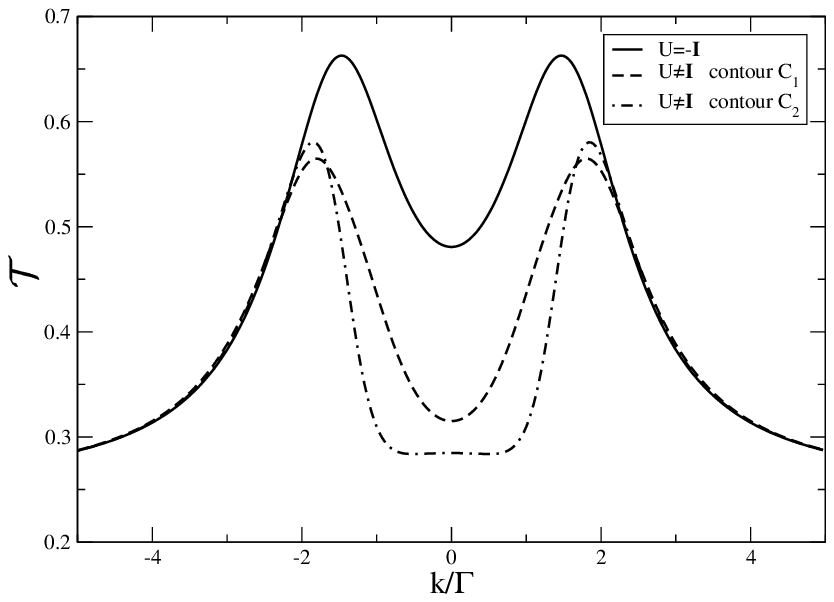}
\caption{\label{nonabelian:fig}
Total transmission through an interferometer 
for a non-Abelian gauge potential~\cite{zhou1}, describing spin-flip 
scattering through a  magnetic atom,  
calculated from Eq (\ref{T:eq}), for different choices of contours, with 
$J/\Gamma=1.5$. The solid line corresponds to the case when 
no geometric phase is present. The dashed lines gives typical results 
after inclusion of the geometric phase. 
There are two peaks at the resonances $k/\Gamma=\pm J/\Gamma$. 
Note that it is impossible to calculate the relative values 
of $t_{ref}$ and 
$t_{QD}$ theoretically, due to the nature of the interferometer. 
As a sample case we take  
the phase difference between $t_{ref}$ and $t_{QD}$ to be $\pi$, 
so that $t_{ref}=-{\mathbf I}$. 
The smaller contour chosen is 
$C_1(\theta_1,\phi_1,\theta_2,\phi_2)=[\pi/2,0,\pi/4,\pi/2]$ 
and the larger is 
$C_2(\theta_1,\phi_1,\theta_2,\phi_2)=[\pi/2,0,\pi/4,\pi]$
(in the same notation as Ref. \cite{zhou1}).
}
\end{figure}

Note that the relative intensity of the two paths for the interferometer 
cannot be calculated theoretically, therefore we choose one specific value 
for the relative intensity. 
Unlike in the Abelian case, the non-Abelian geometric phase, $U$, is gauge 
dependent. Therefore we focus on the transmission. 
In Fig.~\ref{nonabelian:fig} we plot transmission resulting from 
Eq.(\ref{T:eq}) 
as a function of $k/\Gamma$ for a contour
which is a spherical rectangle. 
The parameters $k,J,$ and $\Gamma$ are defined in Ref.~\onlinecite{zhou1}. 
The solid line presents results when the geometric phase is absent. 
After inclusion of the geometric phase the transmission changes significantly 
in both amplitude and shape. Especially, the two peaks at the resonances 
$k/\Gamma=\pm J/\Gamma$ shift due to an energy splitting coming from the 
geometric phase $U$, i.e., during the 
adiabatic change the system moves out of resonance. 

In summary, we developed a theory to describe geometric phases for 
scattering states, and generalized it to the spin-dependent case. 
We have also proposed an experimental setup to directly observe the effect 
from the scattering geometric phase. The effect should be large enough to be 
detected in an open interferometer, and observed as oscillations in the 
current across the device. 

\acknowledgments 
This work was supported by the Australian Research Council.
U.\ Lundin acknowledges the support from the Swedish foundation for
international cooperation in research and higher education (STINT).
We thank Silvio Dahmen, Mark Gould and Jon Links for proof-reading the 
manuscript and discussions.

\begin{thebibliography}{99}

\bibitem{schuster}
R. Schuster {\it et al.}, 
{\it Nature} {\bf 385}, 420 (1997);

\bibitem{ji}
Y. Ji {\it et al.}, 
{\it Science} {\bf 290}, 779 (2000). 

\bibitem{bena}
C. Bena, S. Vishveshwara, L. Balents, and M.P. Fisher, 
{\it Phys.\ Rev.\ Lett.} {\bf 89}, 037901 (2002).

\bibitem{kastner}
M.A. Kastner, 
{\it Rev.\ Mod.\ Phys.} {\bf 64}, 849 (1992).

\bibitem{yacoby}
A. Yacoby, M. Heiblum, D. Mahalu, and H. Shtrikman, 
{\it Phys. Rev. Lett. }  {\bf 74}, 4047 (1995).

\bibitem{berry}
M.V. Berry, 
{\it Proc. Roy. Soc. London, Ser. A} {\bf 392}, 45 (1984). 

\bibitem{jones}
J. Jones, V. Vedral, A.K. Ekert, and C. Castagnoli, 
{\it Nature} {\bf 403}, 869 (2000).

\bibitem{zanardi}
P. Zanardi, and M. Rasetti, 
{\it Phys. Lett. A} {\bf 264}, 94 (1999).

\bibitem{schiff}
L. I. Schiff, {\it Quantum mechanics} (McGraw-Hill, New York, 1955),
p. 290.

\bibitem{sw89}
{\it Geometric Phases in Physics}, edited by 
A. Shapere and F. Wilczek  
(World Scientific, Singapore, 1989).

\bibitem{fazio}
G. Falci {\it et al.}, {\it Nature} 
{\bf 407}, 355 (2000).

\bibitem{zhou1}
H.-Q. Zhou, S.Y. Cho, and R.H. McKenzie, 
{\it Phys. Rev. Lett.} {\bf 91}, 186803 (2003).

\bibitem{b98}
P.W. Brouwer, 
{\it Phys. Rev. B} {\bf 58}, R10135 (1998);
M. B\"uttiker, H. Thomas, and A. Pr\'etre, 
{\it Z. Phys. B} {\bf 94}, 133 (1994);
M. Moskalets and M. B\"uttiker, {\it Phys. Rev. B} 66, 035306 (2002); 
F. Zhou, B. Spivak, and B. Altshuler, 
{\it Phys. Rev. Lett.} {\bf 82}, 608 (1999);
J.E. Avron, A. Elgart, G.M. Graf, and L. Sadun, 
{\it Phys. Rev. Lett.} {\bf 87}, 236601 (2001);
Y. Makhlin, and A.D. Mirlin, 
{\it Phys. Rev. Lett.} {\bf 87}, 276803 (2001);
A. Andreev and A. Kamenev, 
{\it Phys. Rev. Lett.} {\bf 85}, 1294 (2000).

\bibitem{switkes}
M. Switkes, C.M. Marcus, K. Campman, and A.C. Gossard,
{\it Science} {\bf 283}, 1905 (1999). 

\bibitem{marcus}
S.K. Watson, R.M. Potok, C.M. Marcus, and V. Umansky, 
cond-mat/0302492 (unpublished).

\bibitem{msw86}
J. Moody, A. Shapere, and F. Wilczek, 
{\it Phys. Rev. Lett.} {\bf 56}, 893 (1986).

\bibitem{thirring}
H. Narnhofer and W. Thirring, 
{\it Phys. Rev.} {\bf A26}, 3646 (1982).

\bibitem{avron}
J.E. Avron, A. Elgart, G.M. Graf, and L. Sadun, 
{\it J. Math. Phys.} {\bf 43}, 3415 (2002).

\bibitem{zee}
A. Zee, {\it Phys. Rev. A} {\bf 38}, 1 (1988).

\bibitem{bralic}
N.E. Brali\'c, {\it Phys. Rev. D} {\bf 22}, 3090 (1980);
P. M. Fishbane, S. Gasiorowicz, and P. Kaus, {\it Phys. Rev. D}
{\bf 24}, 2324 (1981);
L. Di\'osi, {\it Phys. Rev. D}
{\bf 27}, 2552 (1983).

\end{thebibliography}
\end{document}